\documentclass[twocolumn,showpacs,preprintnumbers,amsmath,amssymb,superscriptaddress]{revtex4-2}
\usepackage{graphicx} 
\usepackage[dvipdfmx]{epsfig} 
\usepackage{bm}
\usepackage{ulem}
\usepackage{amsmath,color}

\begin{document}
\title{Glauber-theory calculations of high-energy nuclear scattering observables using variational Monte Carlo wave functions}
\author{W. Horiuchi}
\email{whoriuchi@omu.ac.jp}
\affiliation{Department of Physics, Osaka Metropolitan University, Osaka 558-8585, Japan}
\affiliation{Nambu Yoichiro Institute of Theoretical and Experimental Physics (NITEP), Osaka Metropolitan University, Osaka 558-8585, Japan}
\affiliation{RIKEN Nishina Center, Wako 351-0198, Japan}

\author{Y. Suzuki}
\affiliation{Department of Physics, Niigata University, Niigata 950-2181, Japan}
\affiliation{RIKEN Nishina Center, Wako 351-0198, Japan}

\author{R.~B. Wiringa}
\affiliation{Physics Division, Argonne National Laboratory, Argonne, Illinois 60439, USA}

\begin{abstract}
Experiments using intermediate- to high-energy radioactive nuclear beams present numerous findings. Extracting important properties of physical observables relies on a firm theoretical analysis. 
Though Glauber theory is believed to work well, no convincing calculation 
has so far been done. We perform Glauber-theory calculations of  both elastic differential cross sections 
and total reaction cross sections for $p+^{12}$C, $^{12}{\rm C}+^{12}$C, and $^6$He+$^{12}$C systems.  
The wave functions of both $^6$He and $^{12}$C are generated by variational Monte Carlo calculations with spatial and spin-isospin correlations induced by realistic two- and three-nucleon potentials. 
Glauber's phase-shift function is computed by Monte Carlo integration up to all orders of nucleon-nucleon multiple scatterings. We show an excellent performance of the Glauber 
description to the selected data on the above systems. We also find that the cumulant expansion of the phase-shift function converges rapidly up to the second order for the above systems. 
This finding will open up interesting applications for the analysis of high-energy nuclear experiments.  
\end{abstract}
\maketitle

{\it Introduction---}Glauber theory~\cite{Glauber} is a versatile theory to deal with high-energy collisions of 
particles appearing in physics. It plays a key role in particle excitations or emissions of a system, see, e.g.,~\cite{Mathur72,Ryckebusch03},
as well as in relativistic hadronic collisions~\cite{Enterria21}. In nuclear physics, 
studies for unstable and exotic nuclei have been done using intermediate to high-energy nuclear beams, and have 
led to discoveries including neutron halos~\cite{Tanihata85}, thick neutron skins~\cite{Suzuki95} and extremely neutron-rich nuclei~\cite{Bagchi20} towards the neutron 
dripline. See also review articles~\cite{Tanihata13, Tanaka24}.

A careful analysis of high-energy scattering observables has been of vital importance to explore the 
properties of those exotic nuclei. 
Though  Glauber's mutiple-scattering theory is expected to be suited for this purpose,  a realistic analysis based on its theory has been a challenge and scarcely been done because  
the computation of Glauber's  phase-shift function (psf) was blocked mainly by two things: One is  a $3\times (A_P+A_T)$-fold  integration  where $A_P$ and $A_T$ are   
the mass numbers of  projectile and target nuclei and 
the other is a physically acceptable way to take into account the breakup effect due to the long-ranged Coulomb potential.  
The former problem was actually resolved in 2002 by the help of a Monte Carlo integration (MCI)~\cite{Varga02}.  Some recipes for the latter problem have been 
used~\cite{NTG, Suzuki03, Horiuchi10, Horiuchi14, Horiuchi16, Horiuchi17}.
We will present a reasonable recipe. 

A $^{12}$C target has often been used in the radioactive beam experiments~\cite{Tanihata85, Tanihata13, Tanaka24}.  
The variational Monte Carlo (VMC) wave functions employed in this work have been generated for the realistic Argonne $v_{18}$ two-nucleon \cite{Wiringa95} and Urbana X three-nucleon potentials \cite{Wiringa14}.  A general review of the VMC method is given in \cite{Carlson15}; the wave function for $^{12}$C has been used successfully in studies of ($p,pN$) reactions \cite{Crespo20,Cravo24}, electron scattering from $^{12}$C \cite{Andreoli24}, and in model studies of neutrinoless double-beta decay \cite{Weiss22}.  A concise description is given in \cite{Weiss22} and one- and two-body densities and momentum distributions are illustrated in \cite{Piarulli23}.
Here we test for the first time 
 Glauber theory unambiguously by comparing to data on total reaction and elastic scattering cross sections. It is also possible to evaluate some approximate ways of computing the psf such as the optical-limit approximation (OLA)~\cite{Glauber,Suzuki03}. We focus on $p+^{12}$C, $^{12}$C$+^{12}$C and $^{6}$He+$^{12}$C cases. See Ref.~\cite{Kaki12} for  the  VMC wave function of $^6$He and  the study of $p+^{6}$He scatterings. More details and comprehensive discussions can be found in the companion paper~\cite{Horiuchi26}. 

{\it Theory---}Two pillars of  Glauber theory are  eikonal and adiabatic approximations~\cite{Glauber,Suzuki03,Bertulani04}.  In order to remove the rapid oscillation of the incident wave, the scattering wave function is cast to $\Psi(\bm{R},\bm{r}_1^P, ... ,\bm{r}_{A_P}^P,\bm{r}_1^T, ... ,\bm{r}_{A_T}^T)
  \equiv  e^{iKZ}\hat{\Psi}(\bm{R},\bm{r}_1^P, ... ,\bm{r}_{A_P}^P,\bm{r}_1^T, ... ,\bm{r}_{A_T}^T)$, 
where $\bm{R}=\bm{R}^P-\bm{R}^T=(\bm{b},Z)$ is the relative coordinate between the two nuclei and $\hbar K$  is the initial momentum of the relative motion. $\bm{b}$ is the two-dimensional impact parameter coordinate.  $\hat{\Psi}$ satisfies the initial  condition,  
$\hat{\Psi} \to \Psi_0^P(\bm{r}_1^P, ... ,\bm{r}_{A_P}^P) \Psi_0^T(\bm{r}_1^T, ... ,\bm{r}_{A_T}^T)$ for $Z\to -\infty$, where $\Psi_0^P$ and $\Psi_0^T$ are the respective normalized ground-state wave functions. 
The spin coordinates are suppressed. The most crucial step is to calculate the psf, $\chi_G(\bm{b})$, defined by
$e^{i\chi_G(\bm{b})}= \left<\Psi_0^P\Psi_0^T\right| \bigl.\hat{\Psi} \bigr>$. 
Each nucleon coordinate $\bm{r}_i$ is expressed as $(\bm{s}_i,z_i)$, 
where $\bm{s}_i$ is the two-dimensional component perpendicular to the beam direction.  
The contribution of the nuclear force to the psf is specified by the 
profile function  $\Gamma_{NN}(\bm{b})=1-e^{i\chi_{NN}(\bm{b})}$ 
determined consistently with nucleon-nucleon ($NN$) scattering data~\cite{Ibrahim08}:  
\begin{align}
  e^{i\chi_N^{\rm tot}(\bm{b})}=\prod_{j=1}^{A_P}\prod_{k=1}^{A_T}[1-\Gamma_{NN}(\bm{b}_{jk})],
\label{profile.eq}
\end{align}
where $\bm{b}_{jk}=\bm{b}+\bm{s}_j^P-\bm{s}_k^T$ is the impact parameter between the $j$-th and $k$-th nucleons. It is too far at present to attempt at deriving $\Gamma_{NN}(\bm{ b})$  from the nuclear force as one needs 
it beyond the hadron production thresholds.
Also, the three-nucleon interaction in such high-energy reaction processes is ignored, since it is strongly suppressed when three nucleons approach each other due to the short-range repulsion of the two-nucleon interaction.
The breakup effect due to the Coulomb potential is accounted for by subtracting the 
phase $\chi_C^{\rm point}(b)=2 \eta \ln{\frac{b}{2D}}$ due to the point-Coulomb potential,  
 $V_C(R)=\frac{Z_PZ_Te^2}{R}$. Here  $Z_Pe$ and $Z_Te$ are the charges of the projectile and target nuclei  and  $\eta=\frac{Z_PZ_Te^2}{\hbar v}$ is the Sommerfeld parameter with the relative velocity $v$.   $D$ is a constant chosen arbitrarily large~\cite{NTG,Suzuki03}.   Extending the 
recipe used in \cite{NTG,Horiuchi16}, we propose 
\begin{align}
  e^{i\Delta \chi_C(\bm{b})}=  
\begin{cases}
\prod_{j=1}^{A_P}\prod_{k=1}^{A_T}
 \left(\frac{|\bm{b}_{jk}|}{b}\right)^{2i\eta \epsilon_j\epsilon_k}    (0\leq b<b_C)  \\
1 \ \ \ \ \ \ \ \ \ \ \ \ \ \ \ \ \ \  \ \ \ \ \ \ \ \ \ \ \ \ \ \  (b_C \leq b).
\end{cases}
\end{align}
$\epsilon_j$ distinguishes proton ($\epsilon_j=1$) or neutron ($\epsilon_j=0$), and $b_C$ is a  cut-off parameter taken to be 
$\sqrt{\frac{5}{3}}(r_p^P+r_p^T)$, where $r_p^P  (r_p^T)$ is the root-mean-square point-proton radius of the projectile (target) nucleus. This choice for $b_C$ ensures that the Coulomb breakup 
occurs in the region where the Coulomb 
potential between the projectile and  target nuclei differs from the point-Coulomb potential. 

The psf $\chi(\bm{b})$ is defined by subtracting $\chi_C^{\rm point}(b)$ from $\chi_G(\bm{b})$:   
\begin{align}
  &e^{i\chi(\bm{b})}\equiv \langle  e^{i\chi_N^{\rm tot}(\bm{b})+i\Delta \chi_C(\bm{b})} \rangle\\
  &=\idotsint d\bm{r}_1^P\dots d\bm{r}_{A_P}^P
  d\bm{r}_1^T\dots d\bm{r}_{A_T}^T\, \notag\\
  &\times\rho^P(\bm{r}_1^P,\dots,\bm{r}_{A_P}^P)\rho^T(\bm{r}_1^T,\dots,\bm{r}_{A_T}^T) e^{i\chi_N^{\rm tot}(\bm{b})+i\Delta \chi_C(\bm{b})}.
\label{fullglauber.eq}
\end{align}
Here,  e.g., $\rho^P(\bm{r}_1^P,\dots,\bm{r}_{A_P}^P)$ denotes $|\Psi_0^P(\bm{r}_1^P,\dots,\bm{r}_{A_P}^P)|^2$, where  the spin coordinates are integrated. 
The elastic scattering amplitude $F(q)$ as a function of the momentum transfer $\bm{q}$
  ($q=|{\bm{q}}|$) is given  by a Fourier transform 
\begin{align}
& F(q)
=e^{-2i\eta \ln (2KD)}\notag\\
& \times\Big\{F_C(q)+\frac{iK}{2\pi}\int d\bm{b}\,
  e^{-i\bm{q}\cdot\bm{b}+2i\eta \ln(Kb)}
  (1-e^{i\chi{(\bm{b})}})
 \Big\},
\label{Glamp.eq}
\end{align}
where $F_C(q)$ is the Rutherford scattering amplitude.  
The elastic differential cross section  reads as 
$  \frac{d\sigma}{d\Omega}=|F(q)|^2$, and 
the total reaction cross section $\sigma_R$ is 
given by integrating the absorption probability with respect to $\bm{b}$, $\sigma_R=\int d\bm{b}\,\left(1-|e^{i\chi(\bm{b})}|^2\right)$. Note that both cross sections are independent of $D$. 

Because both $\rho^P$ and $\rho^T$ are positive-definite, one can perform the multi-dimensional integration in Eq.~(\ref{fullglauber.eq})  by the MCI. 
Its advantage is:  (i) no restriction of the target or projectile wave functions, (ii) 
a full inclusion of the multiple-scattering operator, and (iii) its simplicity~\cite{Varga02}. As far as we know, none but ~\cite{Varga02,Kaki12,Nagahisa18} has applied the MCI to obtain the psf. 
Instead  some approximations have been employed. The accuracy of those approximations 
can be tested by the cumulant expansion~\cite{Glauber, Kubo62, Hufner81, Ogawa92}. See also ~\cite{Bigorda23,Huang23} for recent applications of the cumulant expansion to other physics problems. Because the contribution of $\Delta \chi_C(\bm{b})$ to the psf is found to be very small, we ignore it in  Eq.~(\ref{fullglauber.eq}) and define  
\begin{align}
G(\bm{b},\lambda)&\equiv\Big\langle \prod_{i=1}^{A_P}\prod_{j=1}^{A_T}(1-\lambda \Gamma_{NN}(\bm{b}_{ij})\Big\rangle \nonumber \\ 
&=1\!+\!\sum_{n=1}^{A_PA_T}\mu_n(\bm{b})\lambda^n,
\end{align}
where $\mu_n(\bm{b})$ is the $n$-th moment
\begin{align}
\mu_n(\bm{b})
=(-1)^n \Big\langle \prod_{k=1}^n \Big( \sum_{i_{k}=1}^{A_P}\sum_{j_{k}=1}^{A_T} \Gamma_{NN}(\bm{b}_{i_{k}j_{k}})  \Big) \Big\rangle.
\label{mom-n}
\end{align}
All of $\mu_n(\bm{b})$'s can be calculated by the MCI. 

The psf  is given by $i\chi(\bm{b})=\ln G(\bm{b},1)$. 
A Taylor expansion of $\ln G(\bm{b},\lambda)$ at $\lambda=0$, i.e., the cumulant 
expansion of $G(\bm{b},\lambda)$, reads as  
$\ln G(\bm{b},\lambda)=\sum_{n=1}^{\infty} \frac{1}{n!}\kappa_n(\bm{b}) \lambda^n$. 
The $n$-th cumulant $\kappa_n(\bm{b})$ is expressed in terms of the moments  $\{\mu_i(\bm{b}),  i=1,\cdots, n\}$. E.g., 
\begin{align}
\kappa_1(\bm{b})=\mu_1(\bm{b}),\ \ \ \ \kappa_2(\bm{b})=2\mu_2(\bm{b})-\mu_1(\bm{b})^2.
\label{cum-n}
\end{align}
Substituting $\ln G(\bm{b},1)$ gives the cumulant expansion for the psf:
\begin{align}
& e^{i\chi(\bm{b})}
 =\exp\Big[\mu_1(\bm{b})+\frac{1}{2}\left(2\mu_2(\bm{b})-\mu_1(\bm{b})^2\right)   \Big. \notag\\
   &\quad \Big.+\frac{1}{6}\left(6\mu_3(\bm{b})-6\mu_2(\bm{b})\mu_1(\bm{b})+2\mu_1(\bm{b})^3\right)+\dots\Big].
\end{align}
As seen in Eq.~(\ref{mom-n}), the integral denoted $\langle \cdots \rangle$ 
depends on the properties of the many-particle densities. The higher the order of the cumulant or moment is, the more detailed properties of $\rho^P$ and $\rho^T$ show up, 
that is, the average, the variance, and the skewness, etc. of the many-body densities contribute to the integral.  In what follows, Full means that the psf  is calculated completely including $\Delta \chi_C(\bm{b})$ term by MCI, whereas  cumu-$n$ approximates the psf up to the $n$-th cumulants.

The OLA approximates $e^{i\chi(\bm{b})}$ by $e^{\mu_1(\bm b)}$. 
It has often been employed because $\mu_1(\bm{b})$ is obtained from the one-body densities of the projectile and target nuclei,  The OLA does not work well, however,  
especially for reactions involving 
a halo nucleus~\cite{Bertsch90, Ogawa92, Al-Khalili96a, Al-Khalili96b, Suzuki03, Horiuchi07,Kaki12, Nagahisa18}. This is because the variance of $\rho^P$ and $\rho^T$ contributes significantly~\cite{Ogawa92,Suzuki03}. 
Approximating the full many-body density by a product of the one-body densities and 
including all orders of the multiple-scattering terms was originally suggested by Glauber~\cite{Glauber} 
and has very recently been revisited~\cite{Shabelski21}. With this ansatz, however,  $\chi(\bm{b})$ turns out to be a function of $\mu_1(\bm{b})$:
\begin{align}
e^{i\chi(\bm{b})} \rightarrow \Big(1+\frac{\mu_1(\bm{b})}{A_PA_T}\Big)^{A_PA_T}
 \approx e^{\mu_1(\bm{b})}.
\end{align}
The  resulting psf practically reduces to that of the OLA if $A_PA_T$ is considerably large.  

\begin{figure}[ht]
\begin{center}
  \epsfig{file=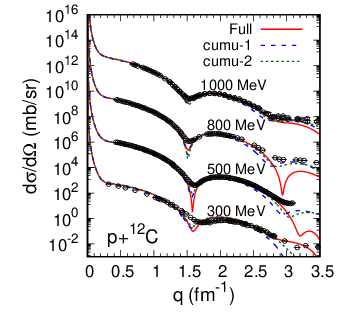, scale=1.4}    
  \caption{$p+^{12}{\rm C}$ elastic differential cross sections at 300, 500, 800, and 1000 MeV as a function       
    of the momentum transfer.   
    The cross sections of three higher incident energies
    are drawn by multiplying a factor 10$^3$ successively. 
    The data are taken from \cite{Meyer85,Okamoto10} (300 MeV), \cite{Hoffmann90} (500 MeV), \cite{Blanpied81} (800 MeV), \cite{Palevsky67,Alkhazov72} (1000 MeV).}
    \label{dcs-C-p.fig}
  \end{center}
\end{figure}

{\it Results---}Figure~\ref{dcs-C-p.fig} displays the $p+^{12}$C elastic differential cross sections.
The full calculation reproduces experiment very well.
  The cumu-1 calculation already gives a nice description up to $q\simeq 2$ fm$^{-1}$. The description of higher $q$ is improved when the second order of the
  cumulant expantion is included.
Though the eikonal approximation imposes $q/K \ll 1$, the calculation works well  up to the second dip. 
For example,  $K$ is $3.7$ fm$^{-1}$ at 300 MeV and the theory reproduces 
the data up to about $q=3$ fm$^{-1}$. 
The Coulomb breakup contribution is small in the $p+^{12}$C scattering.

\begin{figure}[ht]
\begin{center}
  \epsfig{file=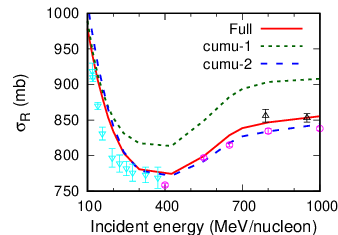, scale=1.4}
  \caption{Energy dependence of the total reaction cross sections of the $^{12}$C$+^{12}$C collision.  
    The data are taken from (inverted triangles)~\cite{Takechi05}  for $\sigma_R$; 
    (circles) \cite{Ponnath24} and  (triangles)  \cite{Ozawa01} for $\sigma_I$.}
    \label{RCS-C.fig}
  \end{center}
\end{figure}

Figure~\ref{RCS-C.fig} shows results of 
$^{12}$C+$^{12}$C total reaction cross sections. Full calculation reproduces the observed cross sections very well above 300 MeV/nucleon. Note that the most recent data on the interaction cross section $\sigma_I$~\cite{Ponnath24} measured above 400 MeV/nucleon have very small error bars.  Since $\sigma_{I}$ is expected to be smaller by dozens of mb than  $\sigma_R$ (See, e.g., \cite{Takechi14,Kohama08}), Glauber theory reproduces experiment excellently.  The cumu-1 approximation overestimates $\sigma_R$ by 30--50 mb beyond 300 MeV/nucleon, while the cumu-2 almost reproduces the full calculation. The adiabatic approximation assumes that  in the center-of-mass system the intrinsic motion of the nucleon in the projectile and target nuclei is much slower 
compared to the  incident energy per nucleon.  The adiabatic assumption 
can be questionable below about 100 MeV/nucleon.  

\begin{figure}[ht]
\begin{center}
  \epsfig{file=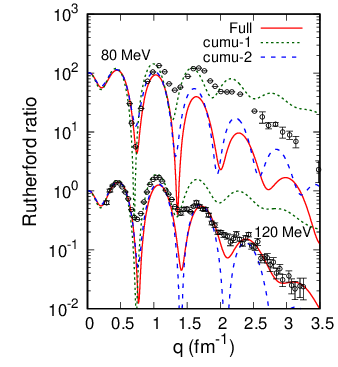, scale=1.4}  
  \caption{Rutherford ratios of the  $^{12}{\rm C}+ ^{12}$C elastic differential 
    cross sections at  80 and 120 MeV/nucleon as the momentum transfer $q$.  
    The cross section at 80 MeV/nucleon is drawn by multiplying a factor 10$^2$.
    The data are taken from ~\cite{Buenerd81} for 86 MeV/nucleon and from 
    \cite{Hostachy87} for 121 MeV/nucleon.}
    \label{dcs-C-C.fig}
  \end{center}
\end{figure}

Figure~\ref{dcs-C-C.fig} plots the $^{12}{\rm C}+^{12}$C elastic differential  cross sections. 
The effect of the Coulomb breakup contribution is small; It is typically less than a few percent at around first and second peak positions at 120 MeV/nucleon. 
 Agreement between theory and experiment in Full calculation is quite satisfactory at 120 MeV/nucleon. At 80 MeV/nucleon the adiabatic assumption is hardly satisfied and the calculation 
reproduces experiment only at $q\lesssim 1$ fm$^{-1}$.  
The cumu-1 approximation overestimates
  the results of the full calculation especially at the backward angles,
  while the cumu-2 result reproduces the full calculation reasonably well, indicating a good convergence in
  the cumulant expansion of the psf.

\begin{figure}[ht]
  \begin{center}
    \epsfig{file=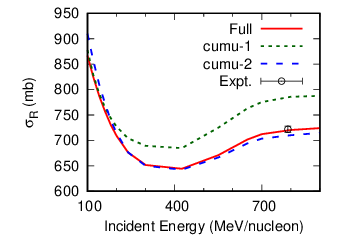, scale=1.4}
  \caption{Energy dependence of the total reaction cross sections of the $^{6}{\rm He}+^{12}$C collision.
     The datum is taken from \cite{Tanihata85b}.
  }
    \label{CS-6He-C.fig}
  \end{center}
\end{figure}

Finally, we show in Fig.~\ref{CS-6He-C.fig} the total reaction cross sections of 
$^{6}$He+$^{12}$C collisions. 
The Full calculation reproduces one interaction cross section datum  very well. 
Since $^{6}$He has no bound excited state, it appears that 
$\sigma_R\approx \sigma_I$ holds in this case very well at 800 MeV/nucleon. 
The OLA or cumu-1 approximation overestimates the cross section considerably.  
The full calculation or the cumu-2 approximation has to be performed to discuss the two-neutron halo 
structure of $^6$He 
based on high-energy scattering data. 

{\it Conclusions---}Describing  high-energy  collisions involving $^{12}$C has played a touchstone 
to extract nuclear properties from experiments using radioactive nuclear beams. 
We have performed Glauber-theory calculations of  both elastic differential cross sections 
and total reaction cross sections for $p+^{12}$C, $^{12}{\rm C}+^{12}$C, and $^6$He+$^{12}$C systems. We have used the VMC wave fucntion for $^{12}$C that is 
now known to 
be most reliable. The phase shift function is computed by Monte Carlo integration that enables us to include all orders of Glauber's multiple-scattering operators with the VMC wave functions. 
The Coulomb breakup effect is accounted for in the profile function by taking into account 
the deviation from the projectile-target point Coulomb potential. 
Within the eikonal and adiabatic approximations of Glauber theory,
all of the experimental data 
are very well reproduced. It is worthy of stressing especially that we reproduce the latest  $^{12}$C$+^{12}$C interaction cross section data accurately measured at 400-1000 MeV/nucleon.

It is an interesting and practically important question whether or not all orders of the multiple-scattering operators have to be taken into account. The question is answered using the 
cumulant expansion of the multiple-scattering operators. We find that taking into account the first and second cumulants gives a fairly good approximation to the complete calculaton. 
This indicates the importance of  the one-body and two-body densities of the projectile and target nuclei in the high-energy collisions, implying
  a connection to the short-range nucleon-nucleon correlations in nuclei (See, e.g., Ref.~\cite{Arrington22} and references therein).
Taking into account the first cumulant is already good enough for a proton-nucleus scattering. 

An important application of  our analysis on the cumulant expansion will be 
the determination of the neutron skin of $^{208}$Pb by $p+^{208}$Pb scatterings. Because of its importance various approaches have been proposed~\cite{Zenihiro10,PREX2, Giacalone23}.  Compared to those, 
a precise measurement of both elastic differential cross sections and total reaction cross sections of $p+^{208}$Pb scatterings above, say, 100 MeV, appears to be quite possible. It is interesting to revisit the analysis of Ref.~\cite{Horiuchi16} by improving the Coulomb breakup 
effect in the profile function as well as a possible bend~\cite{Vitturi87} of  the proton trajectory. The latter effect in fact appears very important in the  $p+^{208}$Pb case, and its 
inclusion can  very easily be done by replacing $Kb$ with $\sqrt{(Kb)^2+{\eta}^2}+\eta$.

\acknowledgments
This work was in part supported by JSPS KAKENHI Grants Nos. 23K22485,
25K07285, 25K01005, and JSPS Bilateral Program No. JPJSBP120247715. 
The work of R.~B.~W. is supported by the U.S. Department of Energy, Office of Science, Office of Nuclear Physics, under Contract No. DE-AC02- 06CH11357; VMC calculations were performed on the parallel computers of the Laboratory Computing Resource Center, Argonne National Laboratory and the Argonne Leadership Computing Facility via the INCITE grant ``Ab initio nuclear structure and nuclear reactions.'' Y.~S. is grateful to M. Kimura of RIKEN for his generosity and encouragement.

\end{document}